# Numerical Contribution of Micro-Pulsed Laser Effect on Copper

**Joseph Dgheim[1], Elio Jad[1], and Rawad Mezher[1]**

[1] Laboratory of Applied Physics (LPA), Group of Mechanical, Thermal & Renewable Energies (GMTER), Lebanese University, Faculty of Sciences II.

E-mail: jdgheim@ul.edu.lb

**Abstract**

A numerical model is developed to study heat, fluid flow and radiation transfers during the interaction between a UV laser beam and copper. Calculations are performed for a laser of Gaussian and Lorentzian shapes of a wavelength of 400nm, a focal spot radius of 50µm and duration of 80µs. In order to describe the transient behaviour in and above the copper target, heat and Navier-Stokes equations are linked to Lambert Beer relationship by taking into account the conduction, and the convection phenomena. The resulting equations are schemed by the finite element method. Comparison with the literature showed qualitative and quantitative agreements for Crater depths and transmission profiles for different laser pulse numbers. Then, the effects of the laser fluences, the Gaussian and Lorentzian shapes on temperature, velocities, melting and evaporation phenomena are studied.

## 1. Introduction

Laser ablation is nowadays used in a growing number of applications, such as chemical analysis and pulsed laser deposition. Despite the many applications, the technique is still poorly understood. Therefore models describing the material evolution in time during short pulse laser irradiation can be helpful in the optimization of the related applications.

During the last twenty years, research in this field has seen greater development regarding new laser sources, mathematical modelling, and industrial applications as reported by Steen (1991) [1]. The theoretical analysis of the laser–material interaction, important for practical applications, is very complex and includes analysis of different physical processes such as material removal, material melting, thermal stresses, shock wave, etc... This prevents successful construction of a general analytical solution; therefore, different numerical procedures have been used in the past by Rozzi et al (1998) [2] and Bianco et al (2004) [3]. The amount of energy transferred from laser radiation to a material's surface is determined by numerous parameters, such as the refractive index, surface roughness, surface coating, laser wavelength, angle of incidence, polarization, focal length, laser power intensity and surface temperature. Experimental results show that for all of the wide band gap materials, the absorptivity increases when laser frequency varies from infrared wavelength to ultraviolet as reported by Pedraza (1998) [4]. The reason was that at ultraviolet wavelength, more energetic photons could be absorbed by a greater number of bound electrons than that at infrared wavelength. The power of the laser has a great influence on absorptivity of material's surface. When the focused power density is not high enough, absorption will be low due to the high reflectivity of the surface. As power density increases, the rate of absorption of individual photons will increase, and thus, the absorptivity will increase remarkably. Beyond a certain power threshold, the absorptivity reaches a constant value or decreases slightly as reported by Cheng et al (2013) [5]. Although there are some empirical and theoretical models of the dependence of the energy coupling efficiency, also called absorptivity, on individual and/or blocks of parameters, to date, there is no comprehensive model that addresses the wide array of parameters that affect energy coupling for laser-material interaction. To address this need, several of the available models are examined here, and a generic methodology is introduced and used to de-couple, classify and re-categorize the parameters. Shankar and Gnamamuthu (1987) [6] have used the finite difference method to solve numerically the transient heat conduction for a moving elliptical Gaussian heat source on a finite dimension solid. They described the heat transfer evolution in iron without taking into consideration the laser power intensity and the refractive index that affect the energy coupling for the laser – material interaction. Lesnic et al (1995) [7] solved the non-linear heat equation for temperature dependent thermal properties, employing the Kirchhoff transformation. In their approach, a constant thermal diffusivity was assumed in order to get a linear heat equation. This provides a limiting case because the thermal conductivity and the heat capacity depend highly on the temperature variations. Jen and Gutierrez (1999) [8] presented an analytical solution for the finite geometry with three different sets of thermal boundary conditions, namely constant wall temperature, insulation and convection at the distant surface. They showed that the size effect plays an important role in determining the temperature distribution, the peak temperature and the location of the peak temperature inside the sample. In their study, they assumed constant thermal properties throughout the domain. However, due to the high irradiation of the localized laser heat source, the thermal properties may change significantly due to the large temperature gradient near the laser heat source. Gutierrez and Araya (2003) [9] carried out the numerical simulation of the temperature distribution generated by a moving laser heat source, by the control volume approach. The numerical model takes into account the radiation and the convection effects. Bianco et al (2004) [10] proposed two mathematical models to evaluate transient conductive fields due to moving laser sources. Melting, evaporation and dopant diffusion are the main mechanisms that result from the laser-matter interaction. As the optical and thermal penetration depths are much smaller than the diameter of the incident laser beam on the surface, the thermal effects described by the heat transfer equation are determined by Hermann et al (2006) [11]. Thus, a model with two thresholds corresponding to the melting threshold of silicon and the ablation threshold of silicon nitride are proposed by Poulain et al (2012) [12]. The authors proposed a simple model that explains the laser ablation and evaluates the size of the ablated area.

Many lasers emit beams that approximate a Gaussian profile. The mathematical function that describes the Gaussian beam is a solution to the paraxial form of the Helmholtz equation. The solution, in the form of a Gaussian function, represents the complex amplitude of the beam's electric field. The behaviour of the field of a Gaussian beam as it propagates is described by a few parameters such as the spot size, the radius of curvature, and the Gouy phase [13]. The Gouy phase which indicates that as a Gaussian beam passes through a focus; it acquires an additional phase shift of $\pi$. In one dimension, the Gaussian function is the probability density function of the normal distribution:

$$G(x) = \frac{1}{\sigma\sqrt{2\pi}} \mathrm{e}^{-(x-x_0)^2/(2\sigma^2)} \quad (1)$$

The Rayleigh range that represents the depth, in which the beam is totally focused, is neglected in our study. The term $\sigma$ that represents the surface in which 63% of the energy distribution is used, and the term $x_0$ that represents the center of the incident laser beam is taken equal to zero.

Other Gaussian profile that varying with time and space is used:

$$\boldsymbol{G(x,t) = \frac{1}{\sigma\sqrt{2\pi}} e^{-(x-x_0)^2/(2\sigma^2)} e^{-(\frac{2t}{\tau})^2}} \quad (2)$$

The timing term in equation (2) has been added, since the laser beam pulse throughout its short pulsation period is not uniform, whereas time increases the intensity decays in an exponential form, this decay is in a Gaussian form. It is also used to have a temporal domain for the incident beam.

Another lasers emit beams that approximate a Lorentzian profile, given by:

$$G(x) = \frac{1}{\pi}\frac{\frac{1}{2}\Gamma}{(x-x_0)^2+(\frac{1}{2}\Gamma)^2} \quad (3)$$

Where, $\Gamma$ is a parameter specifying the width of the laser beam.

In the present work, a copper target initially set at atmospheric pressure and room temperature is irradiated by three types of laser profiles: two types of Gaussian profile while the other type is the Lorentzian one. The effect of the laser beam profiles is very important and influences on the materials surface ablation.

## 2. Mathematical Model

The physical model is formed from a UV laser beam, having pulse duration of 80μs and a wavelength of 400nm, which is used to heat Copper material of $0.5\times1\times1mm^3$ of dimension. The heat transfer occurs in an area $S$ of $0.5\times1mm^2$ (figure 1) during an interval of time in the order of microseconds. The temperature field is computed for the sample, for the border of heated and unheated region. The laser fluence and the time length are chosen to do not treat the phase explosion [14].

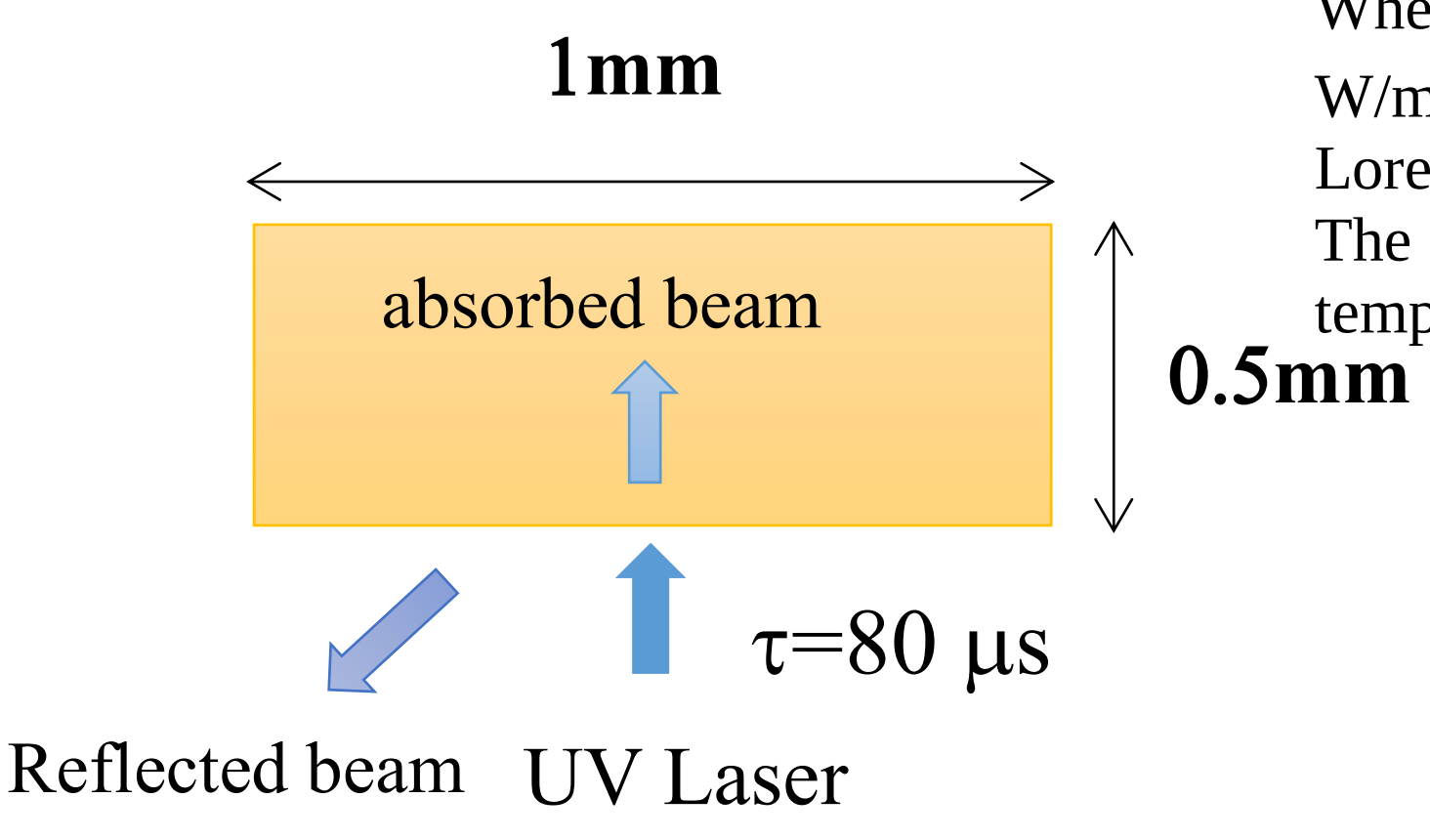

**Figure 1.** Schematic view of laser beam – Copper interaction

The considered solid material has a rectangular shape (figure 2) and is associated to an (*Oxy*) coordinate system, where the origin (*O*) is placed at the center of the system, (*x*) is the curvilinear coordinate, and (*y*) the ordinate counted positively towards the upper side of the system. The thermal-fluid equations are taken without chemical reaction, surface tension, Soret and Dufour effects.

### *2.1 Thermal, fluid flow and radiation equations*

The mathematical model is described by the thermal-fluid equations coupled to the radiation equation of the laser – material interaction. The heat equation is coupled to the Navier-Stokes equations and to the radiation one by the continuity of the conduction, and the convection at the interface. In two dimensional and in the (*Oxy*) coordinates, the thermal fluid equations are the following:

Energy equation:

$$\rho Cp\frac{\partial T}{\partial t} = k\left(\frac{\partial^2 T}{\partial x^2}+\frac{\partial^2 T}{\partial y^2}\right) + Q_{laser} \quad (4)$$

Where $\rho$ is the density, $Cp$ represents the specific heat, $T$ the temperature, $k$ the thermal conductivity and $Q_{laser}$ is the laser beam volumetric power.

The incident laser power is distributed in time and space by Gaussian shapes and in space by Lorentzian one. It can be written according to Lambert-Beer Law [15-16] as the following:

$$Q = nq_0\alpha(1-R)e^{-\alpha y}\times G_{Shape} \quad (5)$$

Where $Q$ is the laser beam in $W/m^3$, $q_0 = E/\tau$ in $W/m^2$ (see table 1), and $G_{shape}$ is the Gaussian or Lorentzian profile.

The volumetric heat flux at the melting temperature is the following:

$$Q_1 = \frac{L_m\rho_m}{\tau} \quad \text{when} \quad T_m <= T < T_e \quad (6)$$

$$Q_2 = \frac{L_v\rho_v}{\tau} \quad \text{when} \quad T >= T_e \quad (7)$$

The final laser beam $Q_{laser}$ becomes:

$$Q_{laser} = Q - Q_1 - Q_2 \quad (8)$$

Continuity equation:

$$\frac{\partial u}{\partial x} + \frac{\partial v}{\partial y} = 0 \quad (9)$$

Where $u$ is the radial velocity and $v$ is the axial velocity.

Momentum equations:

$$\rho\frac{\partial u}{\partial t} = -\frac{\partial P}{\partial x} + \mu\left[\frac{\partial^2 u}{\partial x^2} + \frac{\partial^2 u}{\partial y^2}\right] - F \quad (10)$$

Where $P$ is the pressure and $\mu$ is the dynamic viscosity. The last term of equation (10) is equal to:

$$F = \rho g \beta (T - T_m) \quad (11)$$

Where, $g$ is the terrestrial acceleration and $\beta$ is the thermal expansion coefficient.

The values of these physical parameters are given on table 1.

**Table 1.** Physical parameters of the Copper

| Symbols | Value |
|---|---|
| *E : Micro-Laser Energy* | 5e5 J/m$^2$ |
| *$\tau$ : Laser Pulsation* | 80 µs |
| *n: number of pulses* | 1, 2, 3, 4, ... |
| *$\alpha$: absorption coefficient at wavelength 400nm* | 7.14e7 m$^{-1}$ |
| *R: Reflectivity coefficient* | 0.35 |
| *$L_m$ : Latent heat of fusion* | 133.76 kJ/kg |
| *$L_e$ : Latent heat of evaporation* | 5057.8 kJ/kg |
| *$T_m$: Temperature of melting* | 1356.15 K |
| *$T_e$: Temperature of evaporation* | 2835K |
| *$T_0$ : Ambient temperature* | 293.15 K |
| *$k_s$ : Thermal conductivity (solid state)* | 400 W/(mK) |
| *$k_L$ : Thermal conductivity (liquid state)* | 49.4 W/(mK) |
| *$k_v$ : Thermal conductivity (vapour state)* | 24.6 W/(mK) |
| *$\rho_s$ : Density (solid state)* | 8940 kg/m$^3$ |
| *$\rho_L$ : Density (liquid state)* | 7960 kg/m$^3$ |
| *$\rho_v$: Density (vapour state)* | 6800 kg/m$^3$ |
| *$Cp_s$ : Specific heat (solid state)* | 390 J/(kgK) |
| *$Cp_L$ : Specific heat (liquid state)* | 520 J/(kgK) |
| *$Cp_v$: Specific heat (vapour state)* | 640 J/(kgK) |
| *$\beta$: Thermal expansion coefficient* | 2.5e5 K$^{-1}$ |
| *$k_m$ : thermal conductivity* | $k_s$ for $T<T_m$ or $k_L$ for $T>=T_m$ & $T<T_e$ or $k_v$ for $T>=T_e$ |
| *$\rho_m$ : Density* | $\rho_s$ for $T<T_m$ or $\rho_L$ for $T>=T_m$ & $T<T_e$ or $\rho_v$ for $T>=T_e$ |
| *$Cp_m$ : Specific heat* | $Cp_s$ for $T<T_m$ or $Cp_L$ for $T>=T_m$ & $T<T_e$ or $Cp_v$ for $T>=T_e$ |

### *2.2 Initial and boundary conditions*

- Initial condition for $t<t_0$: (12)

$u$=0

$T$=293.15K

$v$=0

$P$=1atm

- Boundary conditions for $t>t_0$: (13)

<u>At the interfaces</u> (boundary 1): The thermal insulation is applied ($\frac{dT}{dx} = 0$).

Boundary 1: for $x=x_{max}$ $\forall y$

$u=v=0 \quad \frac{\partial T}{\partial x} = 0$

Boundary 1: for $x=-x_{max}$ $\forall y$

$u=v=0 \quad \frac{\partial T}{\partial x} = 0$

<u>At the interfaces</u> (boundary 2): The thermal insulation is applied ($\frac{dT}{dy} = 0$).

Boundary 2: for $y=y_{max}$ $\forall x$

$u=v=0 \quad \frac{\partial T}{\partial y} = 0$

<u>At the surface interface</u> ($y$=0):

At boundary 3 of the figure II.2, the continuity of the conduction and the convection is applied, which is given by the following equation:

$$\vec{n}.(k\nabla T) = h(T_\infty - T)$$

Where $\vec{n}$, is the normal vector.

Boundary 3: for $y$=0 $x>x_{laser}$ & $x<-x_{laser}$

$$u=v=0 \quad -k_s\left.\frac{\partial T}{\partial y}\right|_S = -k_L\left.\frac{\partial T}{\partial y}\right|_L + h\Delta T$$

At boundary 4 of the figure II.2, the continuity of the conduction, the convection and the initial laser fluence is applied, which is given by the following equation:

$$\vec{n}.(k\nabla T) = q_0 + h(T_\infty - T)$$

Boundary 4: For $y$=0 $-x_{laser}<x<x_{laser}$

$$u=v=0 \quad -k_S \left.\frac{\partial T}{\partial y}\right|_S = -k_L \left.\frac{\partial T}{\partial y}\right|_L + h\Delta T + \frac{E}{\tau}$$

At the internal interfaces (boundary 5): The continuity condition is applied.

Far from the interface ( $y \to \infty$):

$$T = T_\infty \text{ and } u=v=0$$

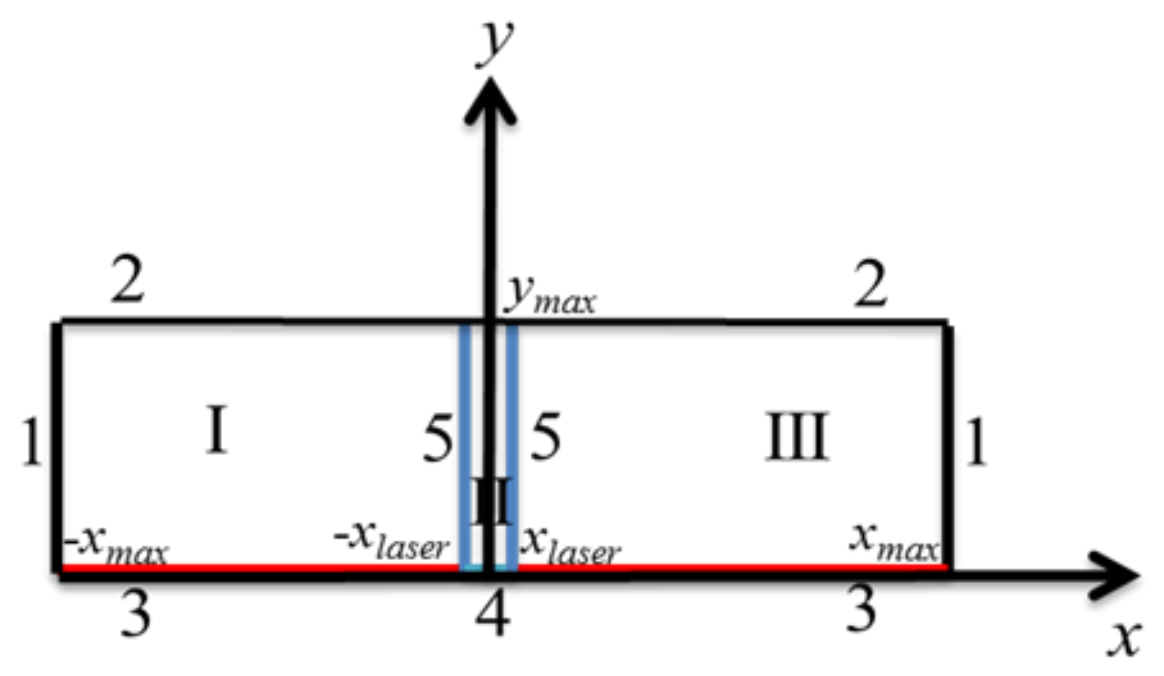

**Figure 2**. Sketch of Copper sample and the model with the boundary conditions

*2.3 Finite element method*

To solve equation (4) by the finite element method, the method of weighed residuals in the Galerkin formulation as reported and detailed by Dgheim (2015 & 2016) [15,16] is used. Using the heat transfer equation and the boundary conditions, the formulation of the weak integral of the thermal problem is obtained as the following:

$$w(T,T^*) = \int_V T^* \rho C_P \dot{T} dV - \int_V T^* \nabla.(k\nabla T) dV - \int_V T^* Q_{laser} dV = 0 \quad (14)$$

The finite element approximation of the equation (14) can be obtained as:

$$w(T,T^*) = \{T^*\}^T ([A]\{\dot{T}\} + [E]\{T\} - \{C\}) = 0 \quad (15)$$

Where, the dot represents differentiation with respect to time. The elementary matrices and the vector of the external heat load are given by:

$$[A] = \int_V [N]^T [\rho Cp][N] dV$$

$$[E] = \int_V [B]^T [K][B] dV + \int_{S\varphi} [N]^T [h][N] dS$$

$$[C] = \int_V [N]^T Q_{laser} dV + \int_{S\varphi} [N]^T (q_0 + hT_\infty) dS$$

[A] is the thermal capacity matrix; [E] is the conduction and the convection matrix and [C] is the nodal flux vector.

By using the finite difference scheme, equation (15), becomes:

$$[A]\frac{\{T\}^{n+1} - \{T\}^n}{\Delta t} + [E]\{T\}^n - \{C\}^n = 0 \quad (16)$$

Equation (16) is solved using an explicit finite difference method and Gauss-Legendre integration.

The same analysis is applied on the continuity and the momentum equations.

## 3. Results and discussion

The variables and constants needed to model this phenomenon are the laser pulse, the absorbed energy, the affected area, and the shape of the pulse (absorbed heat flux distribution).

*3.1 Model Accuracy*

Our mathematical model is tested by reproducing the simulation published by Tesar and Semmar (2008) [17] on iron. Therefore, a rectangular shape for the input energy distribution of the laser beam with a width of 27 ns is considered. The initial temperature of the subdomain was set to 293 K. As a heat source, the energy absorbed on the surface has been used. The surface boundary condition includes the thermal radiation with an iron emissivity equal to 0.1 and heat transfer to ambient room with a convective coefficient ($h$) equal to 10 $W.m^{-2}.K^{-1}$. The scheme of the numerical method is taken not uniform. The mesh elements in the subdomain have a maximal size of 500 nm on the surface (boundaries 3 and 4), but a finer element distribution with a maximal size of 20 nm is used. These mesh sizes are sufficient for the convergence of the numerical model. Smaller sizes don't give more sensitive changes. The time ranged from 0 to 500 ns with a step of 1 ns. The laser beam energy distribution of a rectangular shape is applied on the iron samples in order to study the temperature evolution versus time, for a laser fluence ranged from 500 to 1050 $J.m^{-2}$ (figure 3).

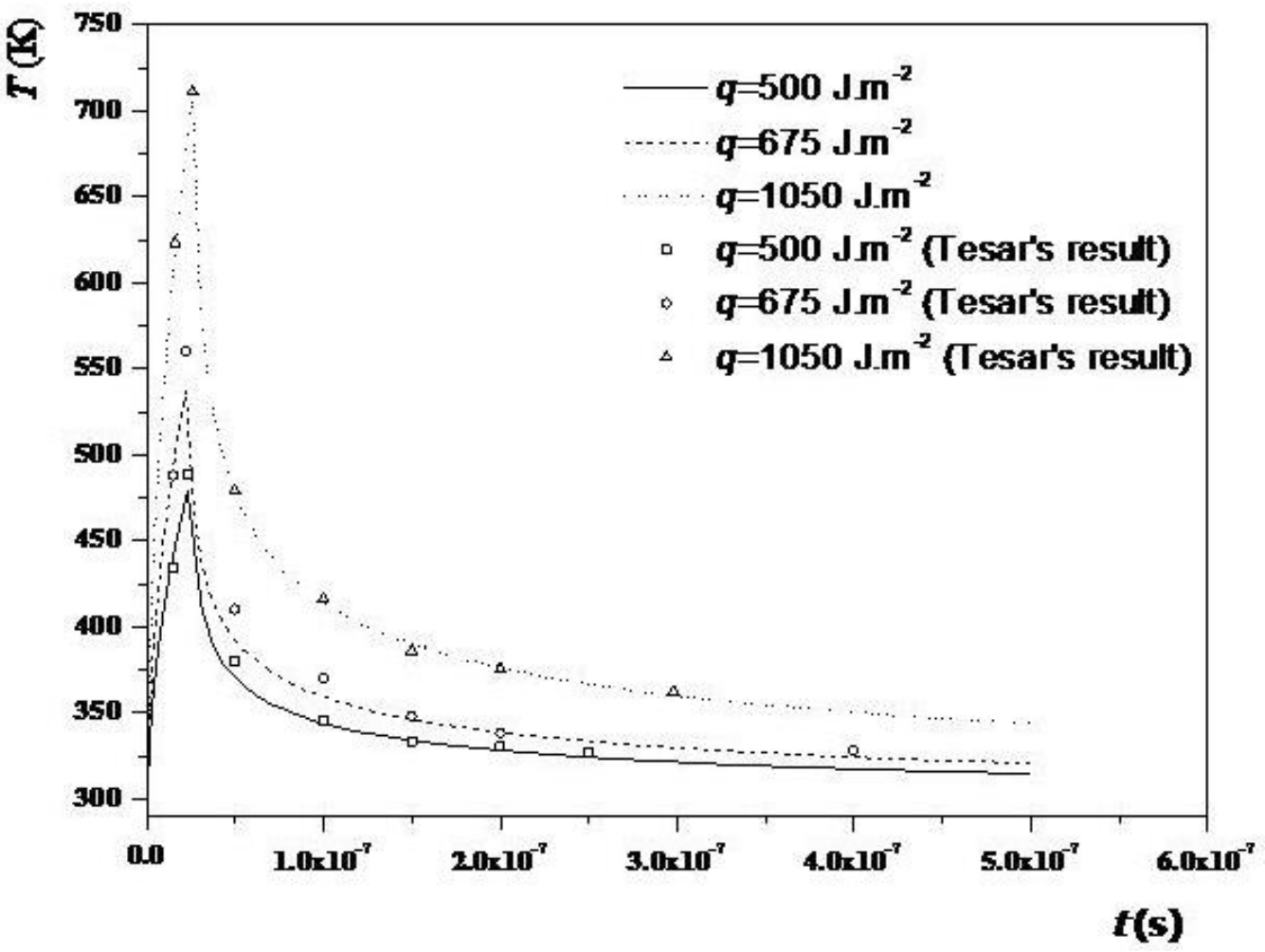

**Figure 3.** Comparison between our numerical results and Tesar et al results.

Figure 3 also shows the comparison between our numerical results concerning the temperature evolution with the results of Tesar and Semmar. Qualitative and quantitative agreements are observed between both results for these laser fluences.

Another accuracy of our results is observed between the Crater diameters for the Gaussian laser pulses for Stainless steel and those calculated by Heinz et al [18]. The following table summarizes these results:

**Table 2.** Comparison between our crater diameters for Stainless steel calculated from our model and Heinz et al experimental results for different pulse numbers

| Symbols | Values | | |
|---|---|---|---|
| Pulse number *n* | 10 | 100 | 1000 |
| Our Model diameter *d* (µm) | 270 | 365 | 680 |
| Diameter calculated experimentally by Heinz et al (µm) | 250 | 340 | ~700 |

The values of our model for Stainless steel crater diameters presented in Table 2 are of very acceptable error as compared to the experimental results of Heinz et al.

*3.2 Copper numerical results*

In the following numerical study, except the laser pulsed duration (27 ns at the Full Width at Half Maximum), the same geometrical, optical and timing parameters in Heintz et al model have been considered in building the Cu model (geometrical shape of the material, affected area of 1×0.5 $mm^2$, microsecond pulsed laser of type UV centred at the wavelength of 400 nm). The model is decomposed in three domains. The mesh is triangular, having 8768 elements in subdomain 1 and 3, and 2304 in subdomain 2. The number of edge elements of our domain is 44. But the number of edge elements when the laser is applied is 144. The time ranges from 0 to 150µs with a step of 5µs. Three types of laser pulses are used. The Gaussian pulses in space and time are used, and the Lorentzian pulse in space. The determined parameters are: the crater diameter (diameter ablated at surface), the ablated mass, and the penetration depth (maximum depth reached by the laser through the medium). The ablation phenomenon will occur when the temperature reaches the evaporation value.

Figure 4 shows the temperature evolution along *x* axis at the surface of the copper (*y*=0), for different laser fluences. The two figures takes into consideration the Gaussian and Lorentzian profiles without time successively. These figures are performed for the same laser of one pulse and the same time of 150µs. When the laser fluence increases the temperature also increases as Gaussian and Lorentzian profiles respectively. Comparison between Gaussian and Lorentzian profiles at a laser fluence of 0.25 $MJ/m^2$ shows clearly the big difference between the peak temperatures of both curves. For the same laser fluence of 0.25 $MJ/m^2$, the temperature of the Lorentzian profile is higher than the one of the Gaussian profile. It reaches 3449 K for the Lorentzian profile and 1177 K for the Gaussian one. By using the laser beam of Lorentzian shape the peak temperature increases quickly to reach the evaporation temperature at the surface of the material.

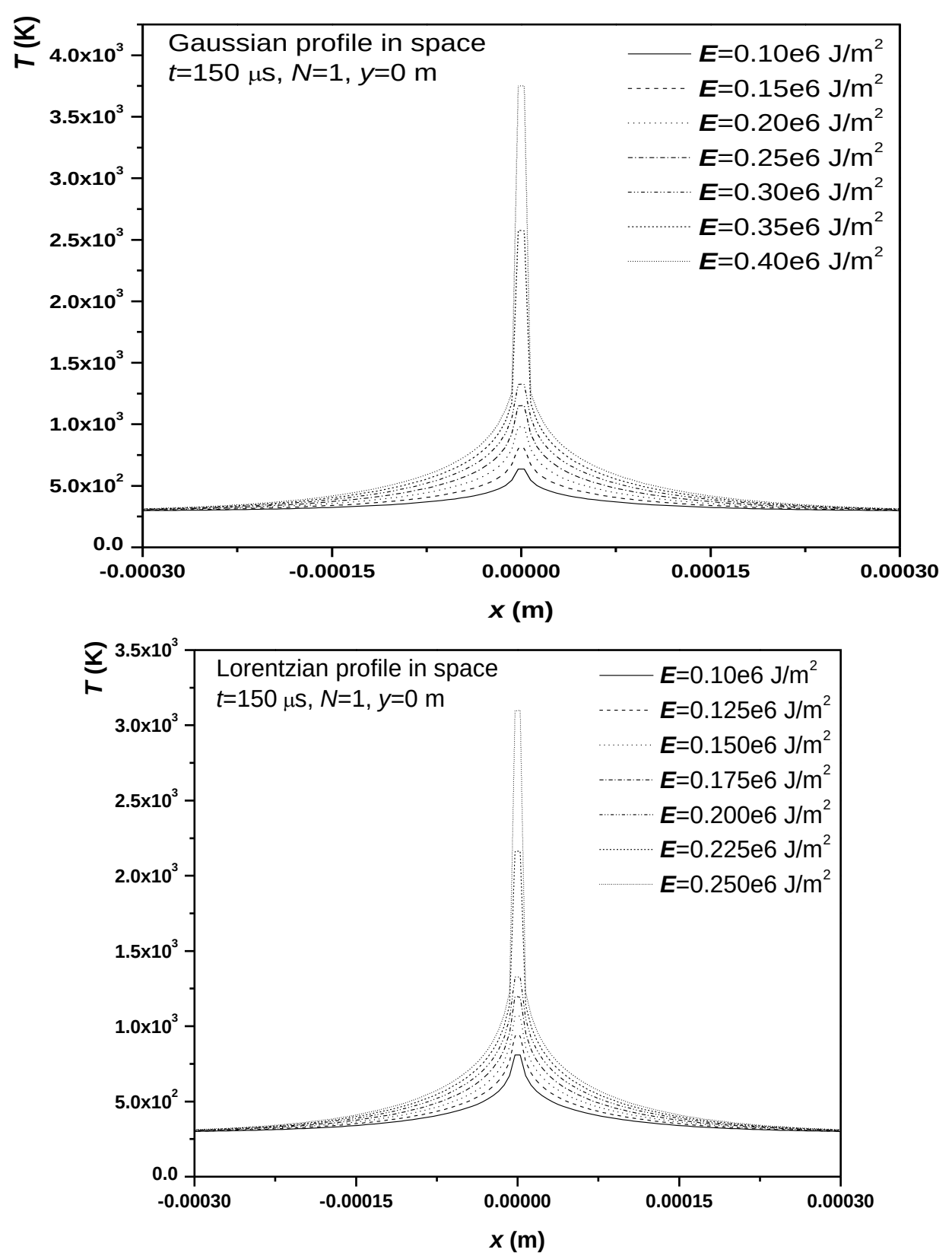

**Figure 4.** Radial temperature evolution for various laser fluences

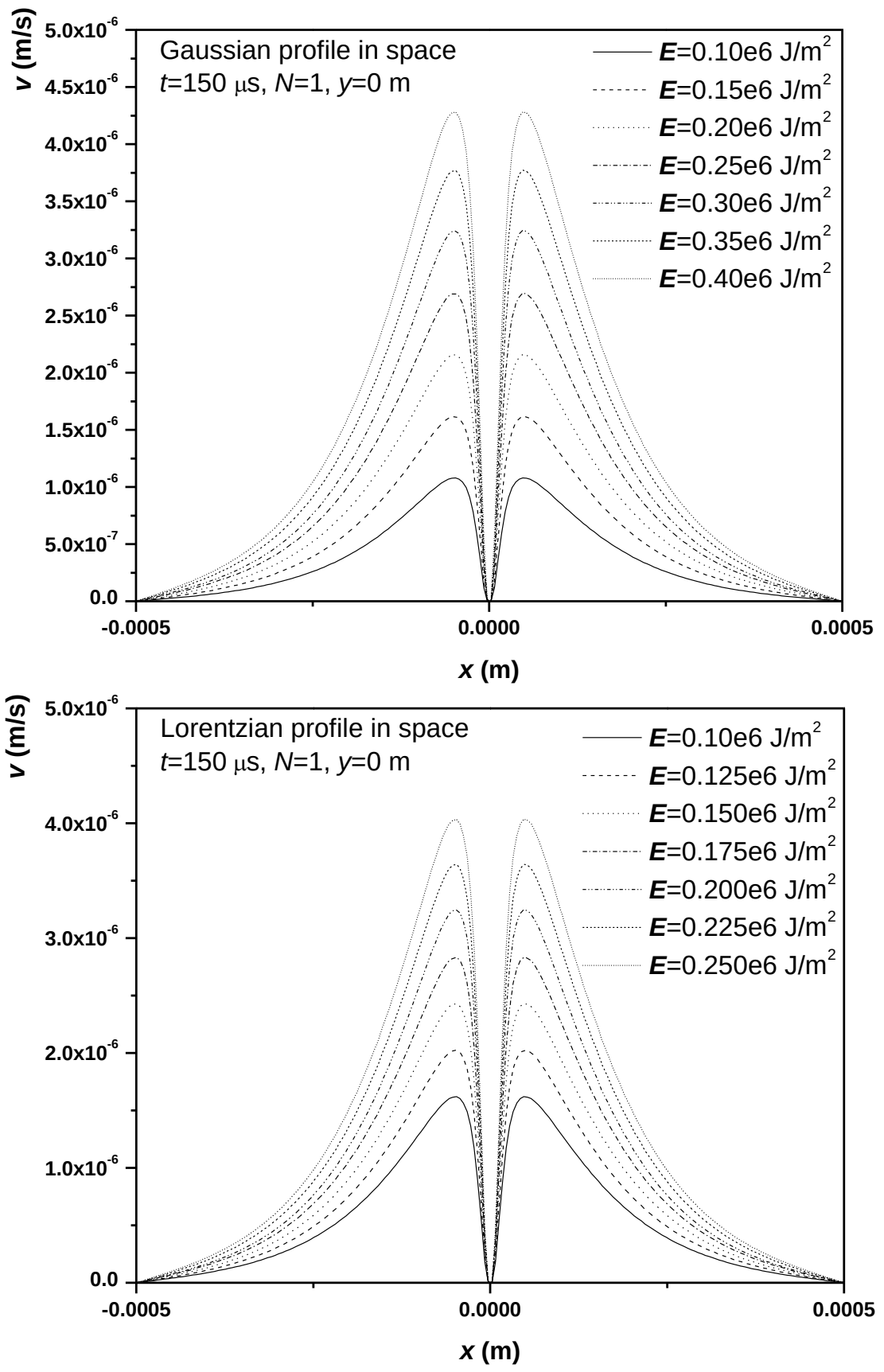

**Figure 5.** Radial velocity evolution for various laser fluences

The radial evolution of the velocity in the crater of the material for various laser fluences is presented in figure 5 for space Gaussian and Lorentzian shapes respectively. At the boundary 4 of the figure 2, the no-slip condition is applied, that explains the zero value of the radial velocity at the center of the material surface. At the material surface, the radial velocity increases from its zero value to reach its maximum and decreases to zero far from the heated region. In fact, this evolution of the radial velocity shows a perturbation in the structure of the Copper surface. When the radial velocity increases this perturbation also increases by presenting the improvement of the convection phenomenon at the surface of the material. The convection phenomenon helps the laser beam to propagate widely into the material by showing a symmetrical velocity profile around the *y* axis. Using Lorentzian profile, the values of the radial velocity increase faster than the ones obtained by Gaussian one. Thus, when the laser fluence increases, the heat transfer is improved and the ablation is realized quickly.

Figure 6 shows the temperature evolution along the depth of the copper material (*y* axis) at a position ($x$=0), for different laser fluences. The two figures takes into consideration the Gaussian and Lorentzian profiles along space respectively. These figures are computed for one laser pulse and for the same time of 150μs. When the laser fluence increases the temperature also increases by using Gaussian profile as well as Lorentzian one. The Temperature of the Lorentzian profile is higher than the one of the Gaussian profile. It reaches 3449 K for the Lorentzian profile and 1177 K for the Gaussian one at laser fluence of 0.25 MJ/m$^2$. The temperature decreases from its maximum value to reach the material ambient temperature far from the surface of the material. This decreasing is explained as the laser beam attenuation by the Copper geometrical structure. When the laser fluence increases, the Copper geometrical structure loses the tendency to resist against this increase in temperature. Comparison between Gaussian and Lorentzian profiles is done at laser fluence of 0.25

MJ/m$^2$. The temperature evolution along Copper depth shows clearly the big difference between both profiles. The resistance of the Copper geometrical structure against the increase in temperature is less important for the Lorentzian shape than the Gaussian one. The increase of the temperature values in the depth of the material (Lorentzian profile) accelerates the ablation phenomenon in the material.

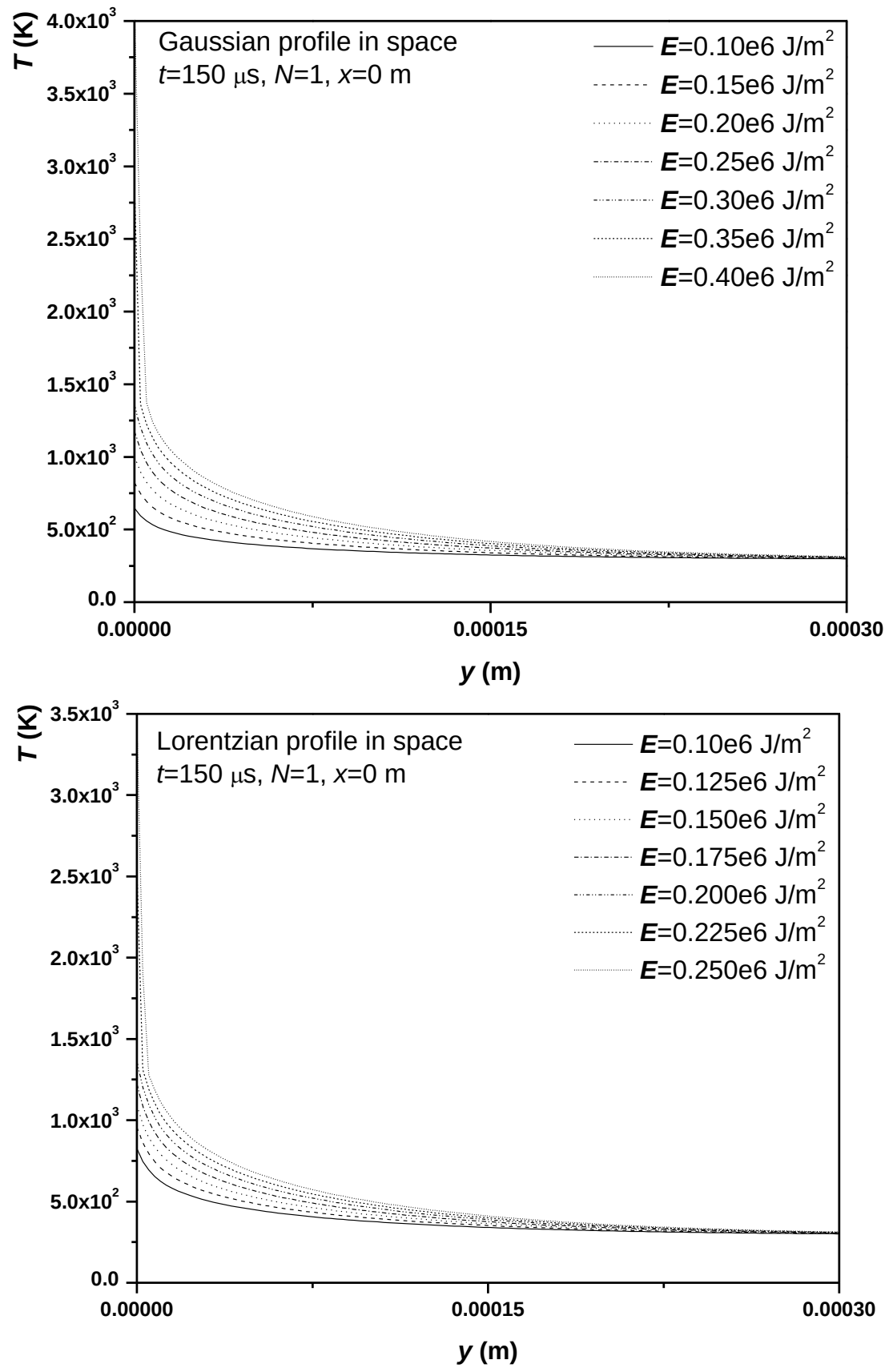

**Figure 6.** Temperature evolution along Copper depth for various laser fluences

Figure 7 shows the temperature evolution along $x$ axis at the surface of the Copper ($y$=0), for different laser pulse numbers. This figure takes into consideration the Gaussian and Lorentzian profiles in space only. These figures are performed at the same laser fluence of 0.35 MJ/m$^2$ and time of 150μs. When the laser pulse number increases the temperature values also increases as Gaussian and Lorentzian shapes respectively. For the same laser pulse number, the temperature values of the Lorentzian profile are higher than the ones of the Gaussian profile. The peak temperature reaches a maximum value of 3789 K for the Lorentzian profile and 1170 K for the Gaussian one after 8 pulses. Thus, the Lorentzian profile takes less pulse numbers to increase the peak temperature to reach the ablated temperature of the material.

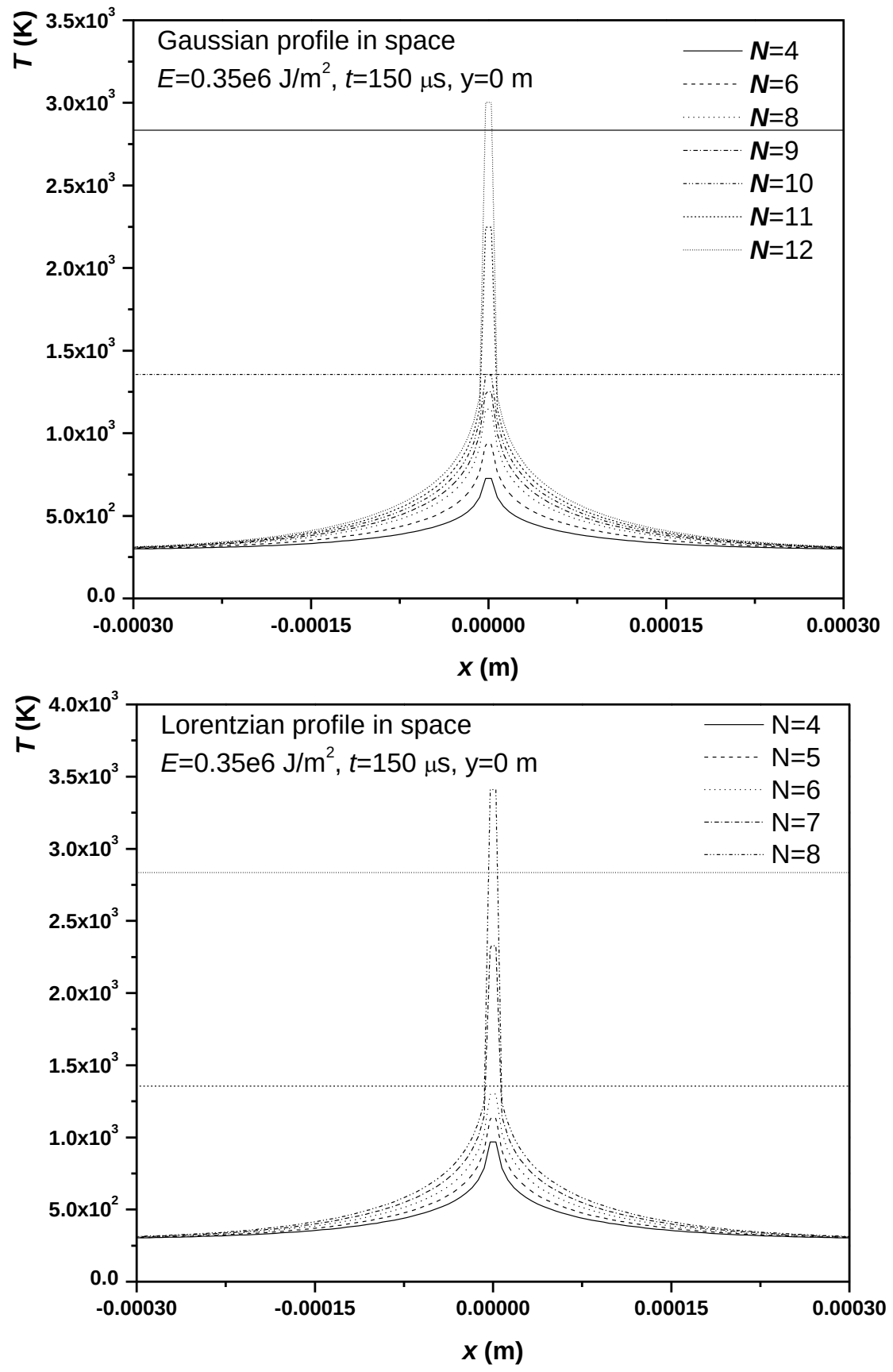

**Figure 7.** Radial temperature evolution for various pulse numbers of space Gaussian shape.

The same study is repeated for the temperature evolution along the depth of the copper sample at a position ($x$=0), for different times (*figure 8*) for space Gaussian and space Lorentzian shapes successively. These figures are performed for the same laser fluence of 0.1 MJ/m$^2$ and for one pulse. When the time increases the temperature values also increase along the depth of the Copper sample. The temperature values of the Lorentzian profile are higher than the ones obtained by the Gaussian profile.

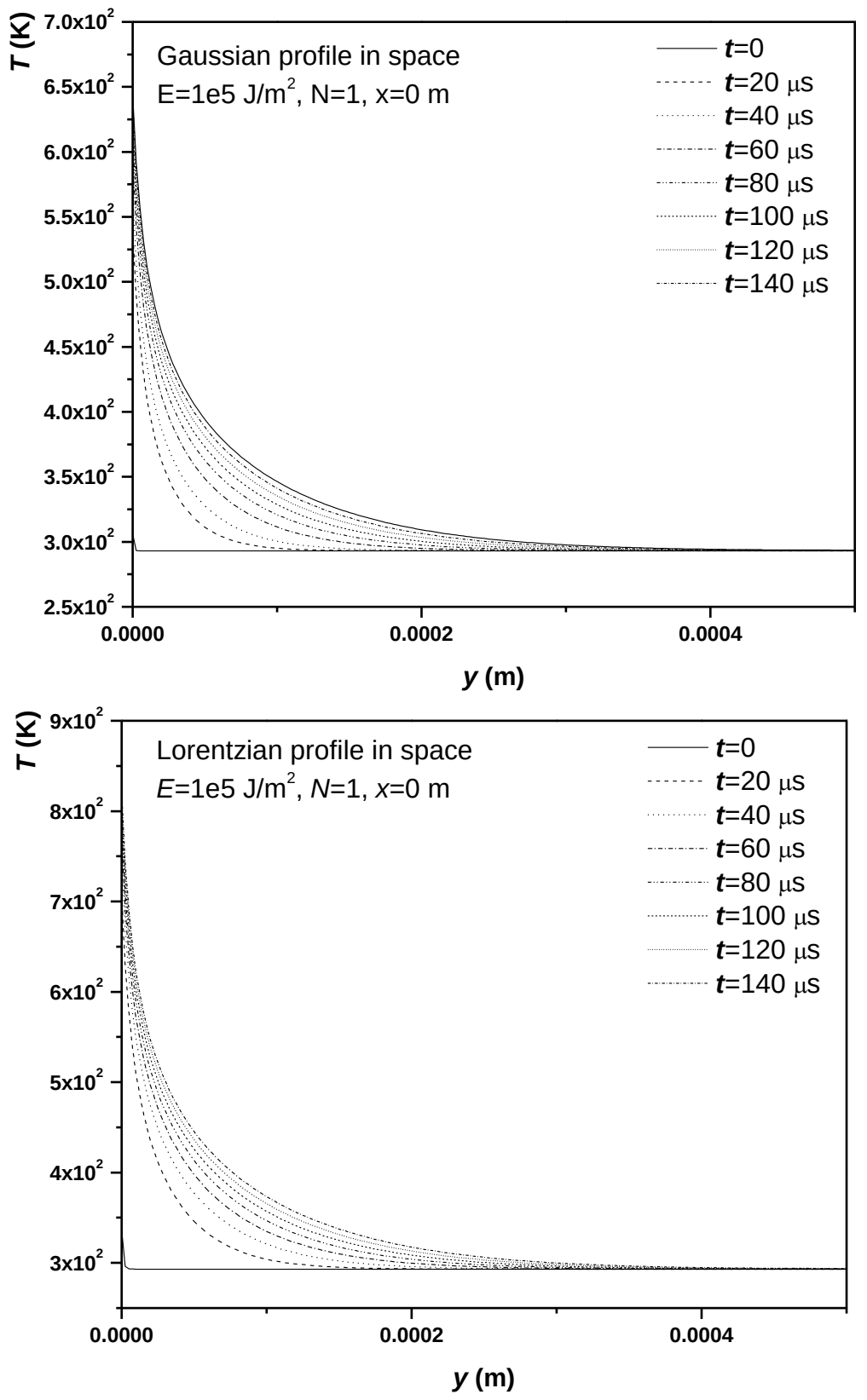

**Figure 8.** Temperature evolution along Copper depth for various times

Comparison between the variation of the temperature evolution versus time for both Gaussian and Lorentzian profiles, for one pulse and the same laser power of 0.1 $MJ/m^2$ is realized. This comparison shows a clear difference between the three curves. For space Lorentzian and Gaussian profiles, the temperature increases from its ambient value to a saturation value after 150µs. The saturation value is the peak value of the temperature evolution for both profiles. The space Lorentzian profile allows us to reach high values of temperatures than the Gaussian ones. For space and time Gaussian profile, the temperature increases rapidly by presenting a peak value and decreases slowly by showing a pulse evolution along time. The peak value of the pulse is observed after 10µs. This Gaussian profile allows us to limit the increasing of the temperature along time and to heat the Copper with a short time (figure 9).

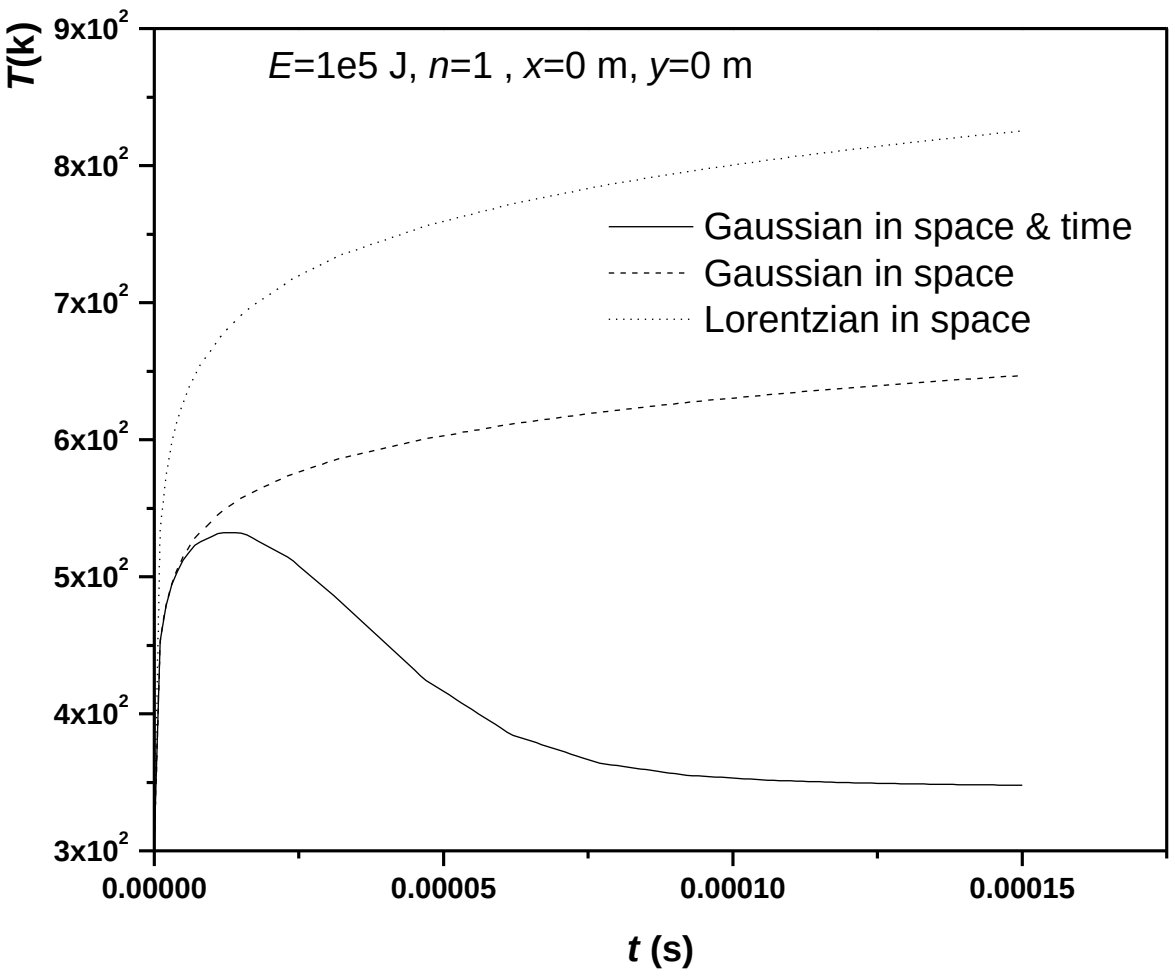

**Figure 9.** Comparison between the temperature evolution versus time of a laser beam of Gaussian and Lorentzian profiles

Comparison between the temperature peak values obtained by both space Gaussian and Lorentzian profiles versus laser fluences, for one pulse and the same time of 150µs, is presented in figure 10. This comparison shows a clear difference between the two curves of both profiles. The Lorentzian profile allows us to reach Copper evaporation temperature faster than the Gaussian one. When increasing the laser fluences, the temperature peak value also increases for both profiles by showing two linear evolutions: the first one corresponds to the heating of the Copper solid state to reach the melting temperature of 1356.15 K, and the second one corresponds to the heating of the Copper liquid state to reach its evaporation temperature of 2835 K. These evolutions are influenced by the thermo-physical and transport properties of the solid, liquid and vapour states. It indicates clearly, the utility of using the Lorentzian profile that can reach faster the ablated temperature of the Copper sample than the Gaussian one, for low laser fluences.

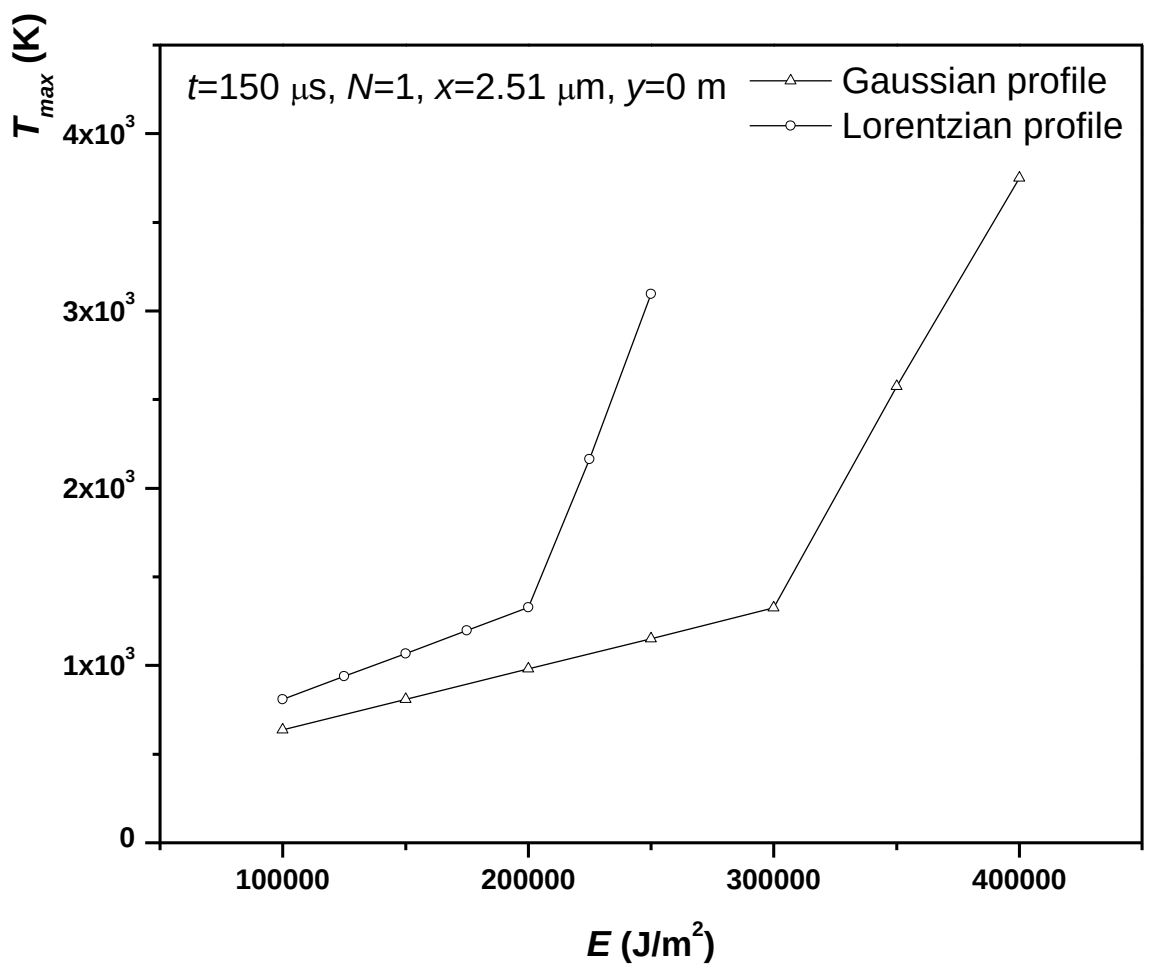

**Figure 10.** Comparison between the peak temperature evolution versus laser fluences (space Gaussian and Lorentzian profiles)

Comparison between the temperature peak values obtained by both space Gaussian and Lorentzian profiles versus laser pulse numbers, for the same laser power of 0.1 MJ/m$^2$, and the same time of 150μs is presented in figure 11. This comparison shows a clear difference between the two curves of both profiles. The Lorentzian profile allows us to reach the ablated temperature faster than the Gaussian one for the same laser pulse number. The same phenomenon as above is observed. It indicates clearly, the advantage of using the Lorentzian profile that can reach faster the ablated temperature of the Copper sample than the Gaussian one, for low laser pulse numbers.

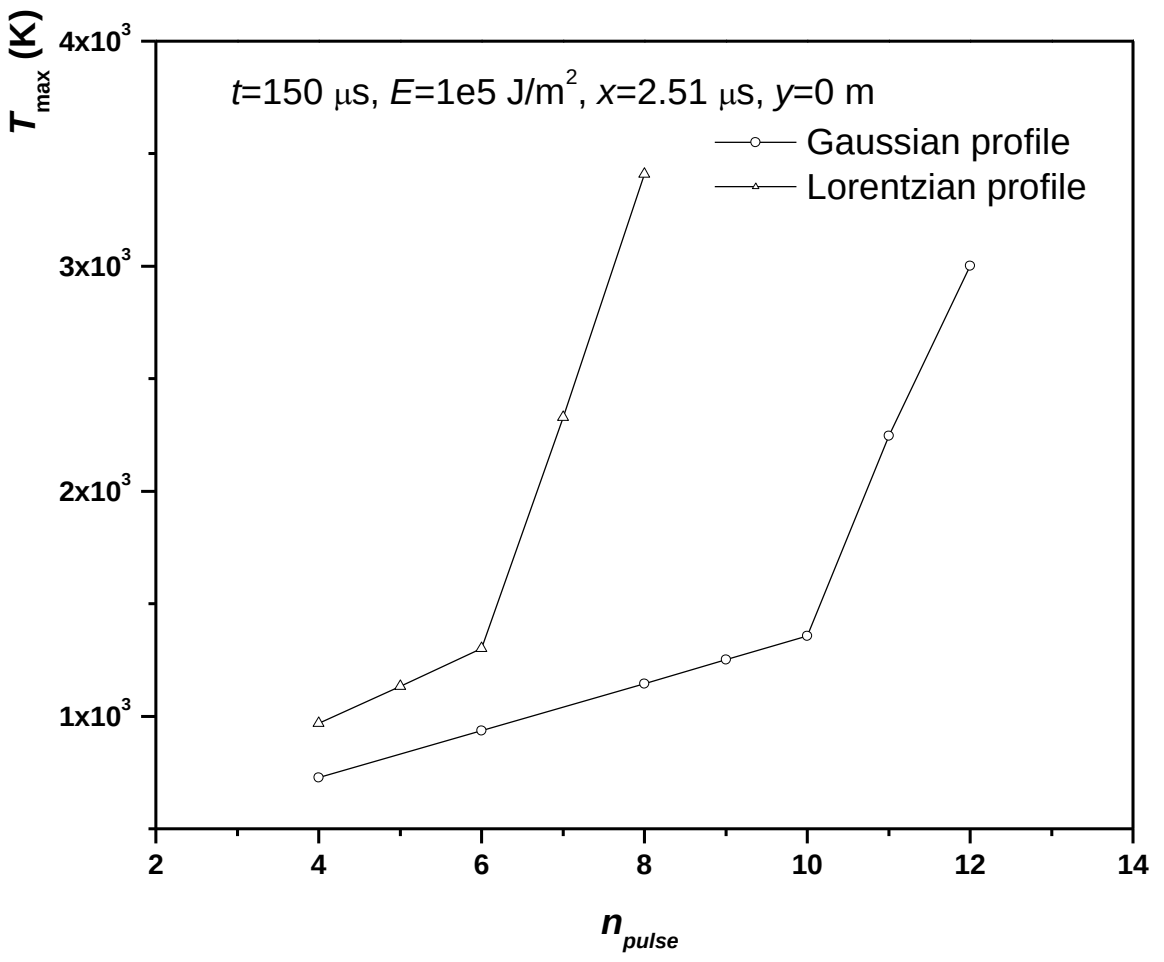

**Figure 11.** Comparison between the peak temperature evolution versus laser fluence for Gaussian and Lorentzian profiles

After calculation of the penetration depths and crater diameters, the ablated mass is determined by assuming a half spherical shape. Then, the formula for calculation of the ablated mass is the following:

$$m_{ablated} = \frac{2}{3}\rho\pi\frac{C_d^3}{8} \tag{17}$$

Where $C_d$ is the crater diameter and $\rho$ is the density of Copper.

Tables 3 and 4 show the ablated dimension, the ablated mass, the crater dimension, and the evaporated mass of the Copper sample for different laser fluences. The ablated dimension and mass increases with the increase of the laser fluences. The crater radius reaches at a fluence value of 0.25 MJ/m$^2$, the value of 0μm for the Gaussian shape and 3.2μm for the Lorentzian shape. The ablated mass reaches at a fluence value of 0.25MJ/m$^2$, the value of 0kg for the Gaussian shape and 9.16E-13kg for the Lorentzian shape.

Tables 5 and 6 show the ablated dimension, and the ablated mass, the crater dimension, and the evaporated mass of the Copper sample for different laser pulse numbers. The ablated dimension and mass increases with the increase of the laser pulse numbers. The ablated radius reaches after 12 pulses, the value of 3.2μm for the Gaussian shape and 9.6μm for the Lorentzian shape. The ablated mass reaches after 12 pulses the value of 9.16E-13kg for the Gaussian shape and 1.26E-11kg for the Lorentzian shape.

**Table 3.** Variation of the mass along different laser fluences (Gaussian shape)

| $q_0$ [J/m$^2$] | Melted radius [μm] | Melted surface [m$^2$] | Melted volume [m$^3$] | Crater radius [μm] | Crater surface [m$^2$] | Crater volume [m$^3$] | Ablated mass [kg] | Evaporated mass [kg] |
|---|---|---|---|---|---|---|---|---|
| 1E05 | 0 | 0 | 0 | 0 | 0 | 0 | 0 | 0 |
| 1.5E05 | 0 | 0 | 0 | 0 | 0 | 0 | 0 | 0 |
| 2E05 | 0 | 0 | 0 | 0 | 0 | 0 | 0 | 0 |
| 2.5E05 | 0 | 0 | 0 | 0 | 0 | 0 | 0 | 0 |
| 3E05 | 0 | 0 | 0 | 0 | 0 | 0 | 0 | 0 |
| 3.5E05 | 5.2 | 8.49E-11 | 4.42E-16 | 0 | 0 | 0 | 2.34E-12 | 0 |
| 3.6E05 | 5.8 | 1.06E-10 | 6.13E-16 | 0 | 0 | 0 | 3.25E-12 | 0 |
| 3.7E05 | 6 | 1.13E-10 | 6.79E-16 | 1.4 | 6.16E-12 | 8.62E-18 | 3.60E-12 | 7.67E-14 |
| 3.8E-5 | 6.4 | 1.29E-10 | 8.24E-16 | 2.13 | 1.43E-11 | 3.04E-17 | 4.37E-12 | 1.38E-13 |
| 3.9E05 | 7.2 | 1.63E-10 | 1.17E-15 | 3.5 | 3.85E-11 | 1.35E-16 | 6.22E-12 | 1.20E-12 |
| 4E05 | 7.73 | 1.88E-10 | 1.45E-15 | 4.26 | 5.07E-11 | 2.43E-16 | 2.34E-12 | 2.16E-12 |

**Table 4.** Variation of the mass along different laser fluences (Lorentzian shape)

| $q_0$ [J/m$^2$] | Melted radius [μm] | Melted surface [m$^2$] | Melted volume [m$^3$] | Crater radius [μm] | Crater surface [m$^2$] | Crater volume [m$^3$] | Ablated mass [kg] | Evaporated mass [kg] |
|---|---|---|---|---|---|---|---|---|
| 1E05 | 0 | 0 | 0 | 0 | 0 | 0 | 0 | 0 |
| 1.25E05 | 0 | 0 | 0 | 0 | 0 | 0 | 0 | 0 |
| 1.5E04 | 0 | 0 | 0 | 0 | 0 | 0 | 0 | 0 |
| 1.75E05 | 0 | 0 | 0 | 0 | 0 | 0 | 0 | 0 |
| 2E05 | 0 | 0 | 0 | 0 | 0 | 0 | 0 | 0 |
| 2.25E05 | 5.3 | 8.82E-11 | 4.68E-16 | 0 | 0 | 0 | 2.48E-12 | 0 |
| 2.35E05 | 5.8 | 1.06E-10 | 6.13e-16 | 0 | 0 | 0 | 3.25E-12 | 0 |
| 2.45E05 | 7.2 | 1.63E-10 | 1.17E-15 | 1.87 | 1.10E-11 | 2.05E-17 | 6.22E-12 | 9.31E-14 |
| 2.5E05 | 7.46 | 1.72E-10 | 1.27e-15 | 3.2 | 3.22e-11 | 1.03e-16 | 6.92E-12 | 9.16E-13 |

.

**Table 5.** Variation of the mass along different pulse numbers (Gaussian shape).

| Pulse Number | Melted Radius [μm] | Melted Surface [m$^2$] | Melted Volume [m$^3$] | Crater Radius [μm] | Crater Surface [m$^2$] | Crater Volume [m$^3$] | Ablated Mass [kg] | Evaporated mass[kg] |
|---|---|---|---|---|---|---|---|---|
| 4 | 0 | 0 | 0 | 0 | 0 | 0 | 0 | 0 |
| 6 | 0 | 0 | 0 | 0 | 0 | 0 | 0 | 0 |
| 8 | 0 | 0 | 0 | 0 | 0 | 0 | 0 | 0 |
| 9 | 0 | 0 | 0 | 0 | 0 | 0 | 0 | 0 |
| 10 | 1.86 | 1.09E-11 | 2.02E-17 | 0 | 0 | 0 | 1.07E-13 | 0 |
| 11 | 5.86 | 1.08E-10 | 6.32E-16 | 0 | 0 | 0 | 3.35E-12 | 0 |
| 12 | 6.93 | 1.51E-10 | 1.05E-15 | 3.2 | 3.22E-11 | 1.03E-16 | 5.55E-12 | 9.16E-13 |

**Table 6.** Variation of the mass along different pulse numbers (Lorentzian shape).

| Pulse number | Melted radius [μm] | Melted surface [$m^2$] | Melted volume [$m^3$] | Crater radius [μm] | Crater surface [$m^2$] | Crater volume [$m^3$] | Ablated mass [kg] | Evaporated mass [kg] |
|---|---|---|---|---|---|---|---|---|
| 4 | 0 | 0 | 0 | 0 | 0 | 0 | 0 | 0 |
| 5 | 0 | 0 | 0 | 0 | 0 | 0 | 0 | 0 |
| 6 | 0 | 0 | 0 | 0 | 0 | 0 | 0 | 0 |
| 7 | 6.13 | 2.53E-07 | 7.20E-11 | 0 | 0 | 0 | 3.84E-12 | 0 |
| 8 | 7.73 | 2.81E-07 | 8.40E-11 | 4.01 | 2.03E-07 | 5.15E-11 | 7.70E-12 | 9.18E-13 |
| 9 | 8.5 | 2.27E-10 | 1.93E-15 | 4.97 | 7.76E-11 | 3.86E-16 | 1.02E-11 | 1.75E-12 |
| 10 | 9.96 | 3.12E-10 | 3.10E-15 | 6.75 | 1.43E-10 | 9.66E-16 | 1.65E-11 | 4.38E-12 |
| 11 | 12.8 | 5.15E-10 | 6.59E-15 | 7.8 | 1.91E-10 | 1.49E-15 | 3.50E-11 | 6.76E-12 |
| 12 | 14.9 | 6.97E-10 | 1.04E-14 | 9.6 | 2.90E-10 | 2.78E-15 | 5.51E-11 | 1.26E-11 |

## 4. Conclusion

Laser ablation on Copper is modeled with the heat and Navier Stokes equations coupled to Lambert-beer's law that takes into consideration two profiles, Gaussian and Lorentzian. The mathematical model is solved by using a finite element scheme of Galerkin formulation.

Comparison between the results of our numerical model and the results of the literature is realized. Qualitative and quantitative agreements are observed between both results. Several parameters that influenced the temperature evolution are discussed. Comparison between the results of our model that takes into consideration Gaussian profile and the one that takes Lorentzian profile, for the laser beam, is done, by changing the laser fluence, the laser pulse numbers, and the time respectively. For the same time of 150μs and for one pulse, when the laser fluence increases the temperature also increases as Gaussian and Lorentzian profiles respectively to reach a higher temperature with Lorentzian pulse. The results are similar, when the laser pulse number increases for the same time of 150μs and the same laser fluence of 0.1 $MJ/m^2$. For the ablation of copper by means of Gaussian and Lorentzian pulses, it was found that the ablation efficiency for the Lorentzian is higher than that of the Gaussian as an ablation in the case of Lorentzian pulses starts from 7 pulses, however a Gaussian has negligible ablation for the same value. The maximum depths of Lorentzian for 12 pulses are considerably higher than those of the Gaussian for the same fluence and under the same pulse number. The ablated mass for Lorentzian profile is higher than that of Gaussian one. Our study shows faster and better results by using a Lorentzian pulse where one can easily reach the ablated heat with lower laser fluence, lower number of pulse and less time.

## 5. Acknowledgements

R. Mezher acknowledges funding from the grant CNRS-L/UL.